\documentstyle[prb,aps,psfig]{revtex}
\input epsf
\begin{document}
\twocolumn[
\hsize\textwidth\columnwidth\hsize\csname@twocolumnfalse\endcsname
\draft
\title{Charge and Spin Response of the Spin--Polarized Electron Gas}
\author{K. S. Yi$^{1,2}$ and J. J. Quinn$^{1,3}$}
\address{$^1$ Department of Physics and Astronomy, University of Tennessee,
Knoxville, Tennessee 37996 \\
  $^2$Department of Physics, Pusan National University, Pusan 609--735,
Korea\\
and\\
$^3$Oak Ridge National Laboratory, Oak Ridge, Tennessee 37831}
\date{\today}
\maketitle
\begin{abstract}
The charge and spin response of a spin--polarized electron gas is investigated
including terms beyond the random phase approximation.
We evaluate the charge response, the longitudinal and transverse spin response,
and the mixed spin--charge response self--consistently in terms
of the susceptibility functions of a non--interacting system.
Exchange--correlation effects between electrons of spin $\sigma$ and
$\sigma^{'}$ are
included following Kukkonen and Overhauser, by using spin--polarization
dependent generalized Hubbard local field factors ${G_\sigma}^{\pm}$ and
${G_{\bar\sigma}}^{\pm}$.
The general condition for charge--density and spin--density--wave excitations
of the system is discussed.
\end{abstract}
]
\setlength{\topmargin}{-2cm}
 
\narrowtext
Response functions relate the induced charge and spin densities to the strength
of an external disturbance and
play an important role in the understanding of many-body systems.
The spin-polarized electron gas (SPEG) is an $n$--electron system with
$n_{\sigma}$ electrons of spin $\sigma$ and $n_{\bar{\sigma}}$ electrons
of spin $\bar{\sigma}$ embedded in a uniform positive charge background.
(The volume of the system is taken to be unity in this work.)
Previous investigations of the response of the spin polarized electron system
were limited in scope.
Some focussed on the paramagnetic response\cite{Vashishta1} or on the
charge and spin--density fluctuations of a ferromagnetic electron gas within
the Hartree--Fock(HF) approximation\cite{Kim}.
Others used the random phase approximation (RPA)\cite{Rajagopal1}, local
spin--density functional theory\cite{Gunnarsson},
or were limited to the infinitesimally polarized electron
liquid\cite{Yarlagadda}.
However, the role of correlations beyond the RPA in the charge--spin response
has never been examined explicitly for the case of arbitrary spin  polarization
$0\le |\zeta| \le 1$, where $\zeta = (n_{\sigma} - n_{\bar{\sigma}}) /
(n_{\sigma} + n_{\bar{\sigma}})$.

The purpose of this paper is to present a treatment of charge and spin response
in a unified way.
The self--consistent effective potential experienced by an electron of spin
$\vec{s}$ is expressed in terms of the charge density fluctuation $\delta n$
and the spin density fluctuation $\delta \vec{m}$.
The exchange--correlation interactions between electrons of the same
spin ($ss$) or of opposite spins ($s\bar{s}$) are included by employing
Hubbard--type spin--dependent local--field factors ${G_s}^{\pm}$ and
${G_{\bar{s}}}^{\pm}$.
The self--consistent linear response method of Kukkonen and Overhauser
\cite{Kukkonen} is extended to a SPEG by generalizing the local field factor.
The charge and spin response to an arbitrary electric and magnetic
disturbance is derived and compared with the existing theories.

We consider an electron gas in the presence of a uniform positive charge
background.
The imbalance in the populations of up and down spins forming a system of SPEG
is caused by an ${\it effective}$ dc magnetic field whose origin needs not be
specified in detail.
Any degree of spin polarization $\zeta$ can be obtained by adjusting the value
of the ${\it effective}$ magnetic field $B$.
We assume the SPEG is disturbed by an infinitesimal external electric
potential ${v_0}^{ext}(\vec{r})$
and magnetic field ${\vec{b}}_0 (\vec{r},t)$.
In response to these external electric and magnetic disturbances, charge and
spin fluctuations are set up in the system, and the Hamiltonian for an electron
with spin $\vec{s}$ can be approximated as

\begin{equation}
H = H_0 + {H_1}^{s},
                                            \label{hamiltonian}
\end{equation}

\noindent
where $H_0$ is the Hamiltonian of a single quasiparticle of the SPEG in the
absence of the external disturbance.
${H_1}^{s}$ is the spin--dependent self--consistent effective perturbation.
The eigenstates and eigenvalues of $H_0$ are given by $|\vec{k},\sigma>$ and
$\varepsilon_{\sigma} (k)$.
In this work we assume the spin--splitting is much greater than the Landau
level splitting and ignore any degree of orbital quantization.
Since the most general disturbance can be decomposed into its Fourier
components, we choose the disturbances ${{v_o}^{ext}}, \vec{b}_0$ and
${H_1}^s$ to vary as $\sim e^{i\omega t - i \vec{q} \cdot \vec{r}}$.
The self--consistent magnetic disturbance $\vec{b}$ is the sum of $\vec{b}_0$
and $4 \pi \vec m$, where $\vec m$ is the induced magnetization.
The Fourier component of the most general ${H_1}^s (\vec{r}, t)$ can be written
as \cite{Zhu1}

\begin{eqnarray}
&&{H_1}^{s} (q, \omega) = \nonumber \\
&&\gamma_0 \vec{s} \cdot \vec{b} +
{v_0}^{ext} +
v(q)[ \delta n \mbox{ } (1-{G_s}^+) - \vec s \cdot \delta \vec{m}  \mbox{ }
{G_s}^- ].
                                            \label{H1}
\end{eqnarray}

\noindent
In Eq.\ (\ref{H1}), for the sake of brevity, the
$\vec{q}$ and $\omega$ dependence of the local fields, fluctuations, and
disturbances has not been displayed.
The parameter $\gamma_0$ is given by $\gamma_0 = \frac {1} {2} g^* \mu_B$ with
$g^*$, $\mu_B$, $\vec{s}$, and $v(q)$
being the effective g--factor, the Bohr magneton, Pauli spin operator, and the
Fourier transform of the bare Coulombic potential, respectively.
Equation\ (\ref{H1}) is the generalization of the effective interaction
Hamiltonian of the SPEG in the presence of infinitesimal magnetic and electric
disturbances.
The local fields ${G_{s}}^+$ and ${G_{s}}^-$ are responsible for charge-- and
spin--induced correlation effects
on an electron of spin $s$\cite{Kukkonen}
\begin{equation}
{G_s}^{\pm} = {G_{ss}}^{xc} \pm {G_{s\bar{s}}}^c,  \mbox{   } \\
{G_{\bar{s}}}^{\pm} = {G_{\bar{s} \bar s}}^{xc} \pm {G_{\bar{s}s}}^c ,
                                            \label{Gpm}
\end{equation}

\noindent
where ${G_{ss}}^{xc}$ (${G_{\bar{s} \bar s}}^{xc}$) and ${G_{s\bar{s}}}^c$
(${G_{\bar{s}s}}^c$) account for the parallel--spin
exchange--correlation and the antiparallel--spin correlation
effects in linear response theory.

The charge-- and spin--density fluctuations $\delta n(q,\omega)$, $\delta
m_i (q,\omega)$ $(i = z, +,$ and $-)$ are given, in terms of the
self-consistent effective perturbation ${H_1}^{s} (q, \omega)$,
using the equation--of--motion of the density matrix\cite{Greene}.

By taking the matrix element of the effective perturbation, Eq.\
(\ref{H1}), with respect to eigenstates $|\vec{k},s>$, then Fourier
transforming the resulting expressions and combining the results with
the definitions of charge-- and spin--density fluctuations, we obtain the
coupled equations for
$<s_1 | {H_1}^{s}(\vec{q}, \omega) |s_2 >$
in terms of the external charge and spin disturbances.
We then solve the coupled equations for $<s_1 | {H_1}^{s}|s_2 >$ and
substitute the corresponding matrix elements back into the expressions of the
fluctuations.
The charge--density fluctuation $\delta n$, longitudinal
spin--density fluctuation $\delta m_z$, and transverse spin--density
fluctuations $\delta m_{+}$ and $\delta m_{-}$ can then be expressed in terms
of a susceptibility matrix $\underline{\chi}$ as

\begin{eqnarray}
 \left( \begin{array}{c}
            -e \delta n     \\
            \gamma_0 \delta m_z
          \end{array} \right)
=
   \left( \begin{array}{cc}
            \chi^{ee} & \chi^{em} \\
            \chi^{me} & {\chi_{\parallel}}^{mm}
          \end{array} \right)
   \left( \begin{array}{c}
            {\varphi_0}^{ext} \\
            {b}_{0z}
           \end{array} \right),	\label{fluct1}
\end{eqnarray}
and
\begin{eqnarray}
\left( \begin{array}{c}
            \gamma_0 \delta m_+  \\
            \gamma_0 \delta m_-
          \end{array} \right)
=
   \left( \begin{array}{cc}
            {\chi_+}^{mm} & 0 \\
            0             & {\chi_-}^{mm}
          \end{array} \right)
   \left( \begin{array}{c}
            {b_0}^+ \\
            {b_0}^-
           \end{array} \right).		\label{fluct2}
\end{eqnarray}

\noindent
Here ${b_0}^{\pm} = b_{0x} \pm i b_{0y}$, and ${\varphi_0}^{ext}$ denotes the
external electric potential corresponding to the
external disturbance ${v_0}^{ext} ( = -e {\varphi_0}^{ext} )$.
The various susceptibilities are written as
\begin{eqnarray}
\chi^{ee} (q, \omega) = \frac {e^2 } {D} [ &&{\Pi_{\sigma \sigma}}^{0} +
{\Pi_{\bar{\sigma} \bar{\sigma}}}^{0} - 16 \pi \gamma_0
{\Pi_{\sigma \sigma}}^0 {\Pi_{\bar{\sigma} \bar{\sigma}}}^0 \nonumber \\
&&+ 2v{\Pi_{\sigma \sigma}}^{0}{\Pi_{\bar{\sigma}
\bar{\sigma}}}^{0} ({G_{\bar{\sigma}}}^- + {G_{\sigma}}^- )
],
                                            \label{chiee}
\end{eqnarray}
\begin{eqnarray}
\chi^{em} (q, \omega) = - \frac {e \gamma_0} {D} [ &&{\Pi_{\sigma \sigma}}^{0}
- {\Pi_{\bar{\sigma} \bar{\sigma}}}^{0} \nonumber \\
&&+ 2v{\Pi_{\sigma \sigma}}^{0}{\Pi_{\bar{\sigma}
\bar{\sigma}}}^{0} ({G_{\bar{\sigma}}}^- - {G_{\sigma}}^- )],
                                            \label{chiem}
\end{eqnarray}
\begin{eqnarray}
\chi^{me} (q, \omega) = - \frac {e\gamma_0 } {D} [ &&{\Pi_{\sigma \sigma}}^{0}
-
{\Pi_{\bar{\sigma} \bar{\sigma}}}^{0} \nonumber \\
&&+ 2v{\Pi_{\sigma \sigma}}^{0}{\Pi_{\bar{\sigma} \bar{\sigma}}}^{0}
 ({G_{\bar{\sigma}}}^+ - {G_{\sigma}}^+ )],
                                            \label{chime}
\end{eqnarray}
\begin{eqnarray}
{\chi_{\parallel}}^{mm} (q, \omega) = \frac {{\gamma_0}^2 } {D} [&&
{\Pi_{\sigma
\sigma}}^{0} + {\Pi_{\bar{\sigma} \bar{\sigma}}}^{0} \nonumber \\
&& - 2v{\Pi_{\sigma \sigma}}^{0}{\Pi_{\bar{\sigma} \bar{\sigma}}}^{0} (2 -
{G_{\bar{\sigma}}}^+ - {G_{\sigma}}^+ )],
                                            \label{chimmz}
\end{eqnarray}
\begin{equation}
{\chi_{+}}^{mm} (q, \omega) = \frac {{\frac{1} {2}} {\gamma_0}^2 {\Pi_{\sigma
\bar{\sigma}}}^{0} }  { 1 + \frac{1}{2}(v {G_{\sigma}}^- - 4 \pi \gamma_0 )
\mbox{  } {\Pi_{\sigma \bar{\sigma}}}^{0} }, 		\label{chimm+}
\end{equation}
and
\begin{equation}
{\chi_{-}}^{mm} (q, \omega) = \frac {{\frac{1} {2}} {\gamma_0}^2
{\Pi_{\bar{\sigma}
\sigma}}^{0} }  { 1 + \frac{1}{2} (v {G_{\bar\sigma}}^- - 4 \pi \gamma_0 )
\mbox{  } {\Pi_{\bar{\sigma} \sigma}}^{0} }.	 \label{chimm-}
\end{equation}
In Eqs.\ (\ref{chiee}) to (\ref{chimm-}),
${\Pi_{s_1 s_2}}^{0}$ is the Lindhard--type electric ($s_1 = s_2$) or spin
($s_1 \ne s_2$) susceptibility.

\begin{equation}
{\Pi_{s_1 s_2}}^{0} (q, \omega)  = \sum_{\vec{k}} \frac {n_0 (\varepsilon_{s_1}
(\vec{k} + \vec{q})) - n_0 (\varepsilon_{s_2} (\vec{k}))} {\varepsilon_{s_1}
(\vec{k} + \vec{q}) - \varepsilon_{s_2} (\vec{k}) - \hbar \omega + i \eta  },
                                            \label{Pi0s1s2}
\end{equation}

\noindent
where $n_0 (\varepsilon_{s} (\vec{k}))$ denotes the equilibrium distribution of
quasiparticles having spin $s$.
The $D$ is given by
\begin{eqnarray}
D (q, \omega) &=&\frac{1}{2} [ 1 - 2 v {\Pi_{\sigma \sigma}}^0 (1 -
{G_{\sigma \sigma}}^{xc}
- {G_{\bar{\sigma} \sigma}}^c )] \nonumber \\
&& \mbox{  } [ 1 + 2 v
{\Pi_{\bar{\sigma} \bar{\sigma}}}^0 ({G_{\bar{\sigma} \bar{\sigma}}}^{xc} -
{G_{\sigma \bar{\sigma}}}^c ) - 8 \pi \gamma_0 {\Pi_{\bar{\sigma}
\bar{\sigma}}}^0 ] \nonumber \\
               &+& \frac{1}{2} [ 1 - 2 v {\Pi_{\bar{\sigma} \bar{\sigma}}}^0
(1 - {G_{\bar{\sigma} \bar{\sigma}}}^{xc}
- {G_{\sigma \bar{\sigma}}}^c )] \nonumber \\
&& \mbox{ } [ 1 + 2 v
{\Pi_{\sigma \sigma}}^0 ({G_{\sigma \sigma}}^{xc} - {G_{\bar{\sigma}
\sigma}}^c)
- 8 \pi \gamma_0 {\Pi_{\sigma \sigma}}^0 ].	\label{D}
\end{eqnarray}
The various terms containing factors proportional to $\pi \gamma_0$ in the
expressions of susceptibilities have their origin in the use of the
self--consistent magnetic disturbance. If we neglect the induced magnetization
$\vec m$, those terms disappear from the expressions for various
susceptibilities.
The spin--polarization dependent Fermi wave number of the majority(minority)
electrons with spin $\sigma (\bar\sigma)$ is given by
\begin{equation}
k_{F_{\sigma(\bar\sigma)}} = k_{F_0} ( 1 \pm \zeta )^{1/3},
							\label{kF}
\end{equation}

\noindent
where $k_{F_0}$ is the Fermi wave number for the unpolarized case.
The expression of ${\Pi_{\bar\sigma \bar\sigma}}^0 $ is
obtained by replacing the quantities of spin indices $\sigma$ in the
expression of ${\Pi_{\sigma \sigma}}^0 $ by that of $\bar\sigma$.
The ${\Pi_{\sigma \bar\sigma}}^0 $ and ${\Pi_{\bar\sigma \sigma}}^0 $ appearing
in Eqs.\ (\ref{chimm+}) and (\ref{chimm-}) are the susceptibility functions of
the spin--flip processes.

In the absence of the perturbation, the charge densities associated with the
majority-- and minority--spin electrons are spatially uniform but unequal.
Hence they have only a non--vanishing $q=0$ Fourier component.
However, in the SPEG a spatially varying electric or magnetic
disturbance with finite wave number $q$
induces electron density fluctuation of each spin,
$\delta n_{\sigma}$ and $\delta n_{\bar{\sigma}}$, and hence a finite spin
density
fluctuation $\delta \vec{m}$.
The ${\chi}^{ee}$, ${\chi_{\parallel}}^{mm}$, and ${\chi_{\pm}}^{mm}$ are the
ordinary charge--, longitudinal spin--, and transverse spin--susceptibilities,
respectively.
The off--diagonal mixed susceptibilities ${\chi}^{em}$(${\chi}^{me}$)
correspond to the charge density response to a magnetic disturbance (the
longitudinal spin density response to an electric disturbance).
The susceptibilities given by Eqs.\ (\ref{chiee}) to (\ref{chimm-})
reduce to appropriate forms for the unpolarized or infinitesimally polarized
limits.
If we set all the local fields ${G_{ss'}} = 0$, we obtain the RPA
susceptibilities of the spin--polarized system.
When $\zeta = 0$, ${\Pi_{\sigma \sigma}}^0 = {\Pi_{\bar{\sigma}
\bar{\sigma}}}^0$ and ${G_{\sigma}}^{\pm} = {G_{\bar{\sigma}}}^{\pm}$.
In this case, $\chi^{ee}$ and ${\chi_{\parallel}}^{mm}$, respectively, reduce
to the well known expressions\cite{Yarlagadda},
\begin{equation}
{\chi^{ee}}_{0} (q, \omega) = \frac{e^2 {\Pi_0}^0 (q, \omega)}{1 - v {\Pi_0}^0
(q, \omega) ( 1 - {G_\sigma}^+ )},
                                        \label{chiee0}
\end{equation}
\begin{equation}
{{\chi_{\parallel}}^{mm}}_{0} (q, \omega) = \frac{{\gamma_0}^2 {\Pi_0}^0 (q,
\omega)}{1 + {\Pi_0}^0 (q, \omega) (v{G_\sigma}^- - 4\pi \gamma_0)},
                                        \label{chimm0}
\end{equation}
where ${\Pi_0}^0 = {\Pi_{\sigma\sigma}}^0 + {\Pi_{\bar\sigma \bar\sigma}}^0 = 2
{\Pi_{\sigma \sigma}}^0$.
The mixed responses vanish in the unpolarized system\cite{Kukkonen}.
But, they become finite and equal to each other in the case of infinitesimally
polarized system\cite{Yarlagadda}, in which ${\Pi_{\sigma \sigma}}^0 \neq
{\Pi_{\bar{\sigma} \bar{\sigma}}}^0$ but ${G_{\sigma}}^{\pm} =
{G_{\bar{\sigma}}}^{\pm}$.
However, since it is not necessary that ${G_{\sigma \bar{\sigma}}}^c =
{G_{\bar{\sigma} \sigma}}^c $ in the system with finite spin polarization, we
conjecture from Eqs.\ (\ref{chiem}) and (\ref{chime}) that, for the most
general case, the charge--spin cross susceptibilities ${\chi}^{em}$ and
${\chi}^{me}$ could be different.
The inequality of the cross--correlation local fields
${G_{\sigma \bar{\sigma}}}^c$, ${G_{\bar{\sigma} \sigma}}^c $ is
expected from the fact that the density $n_{\sigma}^{(h)} (r) $, associated
with the exchange--correlation hole around a given electron with spin $\sigma$
located at the origin, and with $n_{\bar\sigma}^{(h)} (r)$, that around an
electron with spin $\bar\sigma$, are given in terms of the corresponding pair
correlation functions $g_{\sigma \sigma'} (r)$ by
\begin{equation}
n_{\sigma}^{(h)} (r) = n_{\sigma} [ 1- g_{\sigma \sigma} (r)] + n_{\bar\sigma}
[1 - {g_{\sigma \bar\sigma}} (r) ],
							\label{nsigmah+}
\end{equation}
\begin{equation}
n_{\bar\sigma}^{(h)} (r) = n_{\bar\sigma} [ 1- g_{\bar\sigma \bar\sigma} (r)] +
n_{\sigma} [1 - {g_{\bar\sigma \sigma}} (r) ].
							\label{nsigmah-}
\end{equation}
Equations\ (\ref{nsigmah+}) and (\ref{nsigmah-}) show that the
$n_{\sigma}^{(h)} (r) \neq n_{\bar\sigma}^{(h)} (r) $ in the SPEG.
Within the HF or RPA--type approximations,
${G_{\sigma \bar\sigma}}^c = 0 = {G_{\bar\sigma \sigma}}^c$, hence, we have
that ${\chi}^{em} = {\chi}^{me}$ even in the spin polarized
system\cite{Kim,Rajagopal1,Pant}.
In the local spin--density approximation\cite{Gunnarsson}, Gunnarsson
and Lundqvist observed that ${\chi}^{em} = {\chi}^{me}$ by keeping only
the diagonal elements of the matrix $\underline{C}$
in their expressions for the off--diagonal charge--spin
susceptibilities (Eq. (69) in Ref.\cite{Gunnarsson}).
The matrix elements $C_{s,s'}$ in Ref.\cite{Gunnarsson} are directly related
with the local fields ${G_{s(\bar{s})}}^{\pm}$ in the present work.
In the same context, Eqs.\ (\ref{chimm+}) and
(\ref{chimm-}) suggest that, in general, the transverse spin response
functions
${\chi_{+}}^{mm} (q, \omega)$ and ${\chi_{-}}^{mm} (q, \omega)$ could be
different in the SPEG.
Equations\ (\ref{chiee}) to (\ref{chimm-}) can be considered
as definitions of the wave number-- and frequency--dependent local fields
${G_{\sigma}}^{\pm}$ and ${G_{\bar{\sigma}}}^{\pm}$, in terms of the
corresponding fluctuations, in the SPEG.
Within the HF approximation the local fields satisfy the relation
${G_{\sigma}}^+ = {G_{\sigma}}^-$ and, hence, the mixed charge--spin
response functions become equal(${\chi}^{em} = {\chi}^{me}$).

Because the divergences in these response functions give the collective modes
in the system, various susceptibility functions obtained here can be used to
investigate the collective modes such as charge--density and spin--density wave
excitations in the SPEG.
The coupling of charge--density waves and spin--density waves is expected
in the SPEG, and the conditions for the spin--flip transverse modes are
written, from Eqs.\ (\ref{chimm+}) and (\ref{chimm-}), by
\begin{equation}
1+ \frac{1}{2} (v {G_{\sigma}}^{-} - 4 \pi \gamma_0 ) \mbox{  } {\Pi_{\sigma
\bar\sigma}}^{0}  = 0,
\end{equation}
and
\begin{equation}
1+ \frac{1}{2} (v {G_{\bar\sigma}}^{-} - 4 \pi \gamma_0 ) \mbox{  }
{\Pi_{\bar\sigma \sigma}}^{0}  = 0.
\end{equation}
For example,  the divergence of ${\chi_-}^{mm}$ leads us, in a long wavelength
limit, to a mode
\begin{equation}
 \hbar \omega = 2 \gamma_0 B + \frac{n \zeta}{2} (4\pi \gamma_0 - v
{G_{\bar\sigma}}^{-}) + \alpha(\zeta) \mbox{ } {q^2},	\label{trmode}
\end{equation}
where the coefficient $\alpha$ depends, in general, on the degree of spin
polarization of the system and reduces to
$\alpha(1) = \frac{\hbar^2}{2m} (1+\frac{6 \pi \hbar^2}{5m \gamma_0
k_{F_\sigma}})$ for the case of complete spin polarization $\zeta = 1$.
The second term of Eq.\ (\ref{trmode}) disappears as the spin polarization
of the system vanishes.
On the other hand, the general expression for the dispersion relation of the
coupled longitudinal modes is given by the zeros of the  $D(q, \omega)$ defined
by Eq.\ (\ref{D})
\begin{eqnarray}
0 &=& [ 1 - v {\Pi_{\sigma \sigma}}^0 (1 - 2 {G_{\sigma
\sigma}}^{xc} )] [ 1 - v {\Pi_{{\bar\sigma} \bar{\sigma}}}^0 (1 - 2
{G_{\bar{\sigma} \bar{\sigma}}}^{xc} )] \nonumber \\
 &-& 4 \pi \gamma_0 ({\Pi_{\sigma \sigma}}^0 + {\Pi_{{\bar\sigma}
{\bar\sigma}}}^0 ) + 8 \pi \gamma_0 v {\Pi_{\sigma \sigma}}^0
{\Pi_{\bar\sigma \bar\sigma}}^0 (2 -{G_\sigma}^+ -{G_{\bar\sigma}}^+ )
\nonumber \\
 &-&v^2 {\Pi_{\sigma \sigma}}^0 {\Pi_{\bar{\sigma}
\bar{\sigma}}}^0 (1 - 2 {G_{\sigma \bar{\sigma}}}^{c} )(1 - 2 {G_{\bar{\sigma}
\sigma}}^{c} ). 				\label{dispersion}
\end{eqnarray}
In RPA, the above expression reduces to
\begin{eqnarray}
 1 &-& (v+4\pi\gamma_0 )( {\Pi_{\sigma \sigma}}^0 +
{\Pi_{{\bar\sigma}\bar{\sigma}}}^0 ) \nonumber \\
&+& 16 \pi v \gamma_0 {\Pi_{\sigma \sigma}}^0
{\Pi_{{\bar\sigma}{\bar\sigma}}}^0
= 0,
                                             \label{RPAdispersion}
\end{eqnarray}
where ${\Pi_{\sigma\sigma}}^0 \ne {\Pi_{\bar\sigma \bar\sigma}}^0$.
One can expect, from Eq.\ (\ref{RPAdispersion}) coupled modes of
charge--density and spin--density wave excitations,
with a long wavelength limit dispersion relation given by
\begin{eqnarray}
\omega_+ (q) = \Omega_{PL} + \frac{\Omega_{PL}}{2} &[&(1 - \frac{4 n_{\sigma}
n_{\bar\sigma}}{n^2}) \frac{\gamma_0}{e^2} \nonumber \\
&+& \frac{9}{5 n^2}(\frac{{n_{\sigma}}^2}{{q_{TF_{\sigma}}}^2} +
\frac{{n_{\bar\sigma}}^2}{{q_{TF_{\bar\sigma}}}^2})] \mbox{ }q^2 ,
						\label{omega+}
\end{eqnarray}

\begin{equation}
\omega_- (q) = \Omega_{PL} [ \frac{4 n_{\sigma} n_{\bar\sigma}}{n^2}
\frac{\gamma_0}{e^2} -
\frac{9}{5 n^2}(\frac{{n_{\sigma}}^2}{{q_{TF_{\sigma}}}^2} +
\frac{{n_{\bar\sigma}}^2}{{q_{TF_{\bar\sigma}}}^2})]^{1/2} \mbox{ } q.
						\label{omega-}
\end{equation}
Here $\Omega_{PL}$ is the plasma frequency corresponding to the total electron
density $n = n_\sigma + n_{\bar\sigma}$ and $q_{TF_{\sigma(\bar\sigma)}}$ is
the Thomas--Fermi wavenumber\cite{Fetter} of the majority(minority) electrons
of spin $\sigma(\bar\sigma)$.
The terms involving $\gamma_0$ have their origin in the self--consistent
magnetic responses
and, especially, those terms containing $4 n_{\sigma} n_{\bar\sigma}$ result
from the coupling of the electric and magnetic responses in Eq.\
(\ref{RPAdispersion}).
The terms of ${n_\sigma}^2$ and ${n_{\bar\sigma}}^2$ are due to the
contribution to the noninteracting susceptibility of the extra kinetic energy
in the SPEG Fermi sea.
The minus sign in Eq.\ (\ref{omega-}) indicates that, in the long
wavelength region, the slope of the dispersion of the mode $omega_-$ is
suppressed in the presence of the coupling between the oscillations of charge
density and spin density of the SPEG.
When $\zeta = 0$, Eq.\ (\ref{dispersion}) reduces as
\begin{equation}
[ 1 - v {\Pi_0}^0 (1 - {G_{\sigma}}^{+} )] [ 1 + {\Pi_0}^0 ( v {G_{\sigma}}^{-}
- 4 \pi \gamma_0 )] = 0,
                                             \label{disp0}
\end{equation}
which is the product of the conditions of self--sustaining oscillations of
charge and spin densities in a spin--unpolarized system given by Eqs.\
(\ref{chiee0}) and (\ref{chimm0}).
For the case ${G_{\sigma}}^{\pm} = 0$, Eq.\ (\ref{disp0}) reduces to
$ [ 1 - v {\Pi_0}^0 ] [ 1 - 4 \pi \gamma_0 {\Pi_0}^0 ] = 0$, which is the
RPA result of a unpolarized system.
The first factor leads us to the well--known charge density wave excitation due
to Coulomb interaction\cite{Fetter};
$\omega = \Omega_{PL} ( 1 + \frac{9}{10} \frac{q^2}{{q_{TF}}^2})$.
One the other hand, the second factor gives us the spin density wave
excitation, in response to the self--consistent magnetic disturbance, of the
spin--unpolarized Fermi sea;
$\omega = \Omega_{PL}\sqrt{\frac{\gamma_0}{e^2}}( 1 + \frac{9}{5}
\frac{e^2}{\gamma_0} \frac{1}{{q_{TF}}^2})^{1/2} \mbox{ } q$ .
For the case of complete spin--polarization $\zeta = 1$, the frequency and wave
number dependence of various longitudinal susceptibilities given by Eqs.\
(\ref{chiee}) to (\ref{chimmz}) becomes the same and the condition for the
longitudinal collective modes is given by
\begin{equation}
 1 -  {\Pi_{\sigma \sigma}}^0 [ v (1 - 2 {G_{\sigma}}^{xc})  + 4 \pi \gamma_0 ]
= 0 .
                                             \label{disp1}
\end{equation}
If we set ${\Pi_{\sigma \sigma}}^0 \neq {\Pi_{\bar{\sigma} \bar{\sigma}}}^0$
but ${G_{\sigma}}^{\pm} = {G_{\bar{\sigma}}}^{\pm}$,
Eq.\ (\ref{dispersion}) becomes
\begin{eqnarray}
0 &=& 1 - v ({\Pi_{\sigma \sigma}}^0 + {\Pi_{\bar\sigma \bar\sigma}}^0 )
(1 - {G_{\sigma}}^{+} ) + [ {\Pi_{\sigma\sigma}}^0 + {\Pi_{{\bar\sigma}
\bar{\sigma}}}^0 \nonumber \\
&& \mbox{  } \mbox{  }- 4 v {\Pi_{\sigma \sigma}}^0 {\Pi_{\bar{\sigma}
\bar{\sigma}}}^0
(1 - {G_{\sigma}}^{+} )] ( v {G_{\sigma}}^{-} - 4 \pi \gamma_0 ).
                                             \label{dispinf}
\end{eqnarray}
Taking the external magnetic disturbance ${\vec b}_0$ as our effective
magnetic disturbance in Eq.\ (\ref{H1}), instead of the self--consistent field
$\vec b$, makes the factor $4 \pi \gamma_0$ on the right hand side of Eq.\
(\ref{dispinf}) disappears, and the expression reduces to the result of an
infinitesimally spin polarized system\cite{Yarlagadda}.

In summary, a unified treatment of the response of the spin polarized electron
gas is presented in this paper and general expressions for various
susceptibility functions are derived.
The present results reproduce exactly the known results for several
simple situations.
Spin--polarization dependence of the HF local fields is displayed explicitly.
We believe that our results could be useful in understanding electric,
magnetic, and optical properties of a number of spin polarized systems like
ferromagnetics and diluted magnetic semiconductors.

\acknowledgements

This work was supported in part by the Oak Ridge National Laboratory, managed
by Lockheed Martin Energy Research Corp. for the US Department of Energy under
the contract No. DE--AC05--96OR22464.
One of the authors(K.S.Y.) appreciates the supports by the 1995 program of the
Korea
Research Foundation and in part by the BSRI--96--2412 program of the Ministry
of Education, Korea.
The authors would like to thank Drs. J. Cooke, G. F. Giuliani, A. W.
Overhauser, and P. Vashishta for helpful comments.

\end{document}